\documentclass[a4paper,12pt]{iopart}
\usepackage{iopams}
\usepackage{graphicx}

\newcommand{\dd}{\mathrm{d}}

\newcommand{\eq}[1]{(\ref{#1})}

\newcommand{\dotcross}{ \raise 0.65ex\hbox{${\scriptstyle {{_{\displaystyle \cdot}}\atop\times}}$} }
\newcommand{\crossdot}{ \raise 0.5ex\hbox{${\scriptstyle {{_\times}\atop{\displaystyle \cdot}}}$} }

\newcommand{\sumsig}{ \raise -1.3ex\hbox{${{\displaystyle \sum}\atop{\scriptstyle \sigma}}$} }
\newcounter{appnumb}

\begin{document}

\title{Less constrained omnigeneous stellarators}

\author{Felix I. Parra$^{1,2}$}
\eads{\mailto{f.parradiaz1@physics.ox.ac.uk}}

\author{Iv\'an Calvo$^{1,3}$}
\eads{\mailto{ivan.calvo@ciemat.es}}

\author{Per Helander$^4$}
\eads{\mailto{per.helander@ipp.mpg.de}}

\author{Matt Landreman$^5$}
\eads{\mailto{matt.landreman@gmail.com}}

\vspace{0.7cm}

\address{$^1$Rudolf Peierls Centre for Theoretical Physics, University of Oxford, Oxford, OX1 3NP, UK}
\address{$^2$Culham Centre for Fusion Energy, Abingdon, OX14 3DB, UK}
\address{$^3$Laboratorio Nacional de Fusi\'on, CIEMAT, 28040 Madrid, Spain}
\address{$^4$Max-Planck-Institut f\"ur Plasmaphysik, 17491 Greifswald, Germany}
\address{$^5$University of Maryland, College Park, MD 20742, USA}

\begin{abstract}
A stellarator is said to be omnigeneous if all particles have vanishing average radial drifts. In omnigeneous stellarators, particles are perfectly confined in the absence of turbulence and collisions, whereas in non-omnigeneous configurations, particle can drift large radial distances. One of the consequences of omnigeneity is that the unfavorable inverse scaling with collisionality of the stellarator neoclassical fluxes disappears. In the pioneering and influential article [Cary~J~R and Shasharina~S~G 1997 {\it Phys. Plasmas} {\bf 4} 3323], the conditions that the magnetic field of a stellarator must satisfy to be omnigeneous are derived. However, reference [Cary~J~R and Shasharina~S~G 1997 {\it Phys. Plasmas} {\bf 4} 3323] only considered omnigeneous stellarators in which all the minima of the magnetic field strength on a flux surface must have the same value. The same is assumed for the maxima. We show that omnigenenous magnetic fields can have local minima and maxima with different values. Thus, the parameter space in which omnigeneous stellarators are possible is larger than previously expected. The analysis presented in this article is only valid for orbits with vanishing radial width, and in principle it is not applicable to energetic particles. However, one would expect that improving neoclassical confinement would improve energetic particle confinement.
\end{abstract}


\maketitle

%%%%%%%%%%%%%%%%%%%%%%%%%%%%%%%%%%%%%%%%%%%%%%%%%%%%%%%%%%%%%
\section{Introduction}
%%%%%%%%%%%%%%%%%%%%%%%%%%%%%%%%%%%%%%%%%%%%%%%%%%%%%%%%%%%%%

Charged particles are not necessarily confined in the three dimensional stellarator magnetic fields. Particles can move long distances away from the flux surface in which they started, causing large neoclassical transport. For this reason, stellarators must be optimized using sophisticated codes \cite{gori99, subbotin06, kasilov13} to reduce neoclassical transport to acceptable levels \cite{wobig93, nuhrenberg95, anderson95, neilson02}. The pioneering work of Cary and Shasharina \cite{cary97a, cary97b} gave the conditions that the magnetic field on a flux surface has to satisfy to reduce the average particle radial drift to zero. A flux surface that satisfies the conditions given in \cite{cary97a, cary97b} is said to be omnigeneous. It will have neoclassical particle and energy fluxes comparable to those in a tokamak and consequently, negligible compared to turbulent fluxes.

\begin{figure}

\begin{center}
\includegraphics[width = 12 cm]{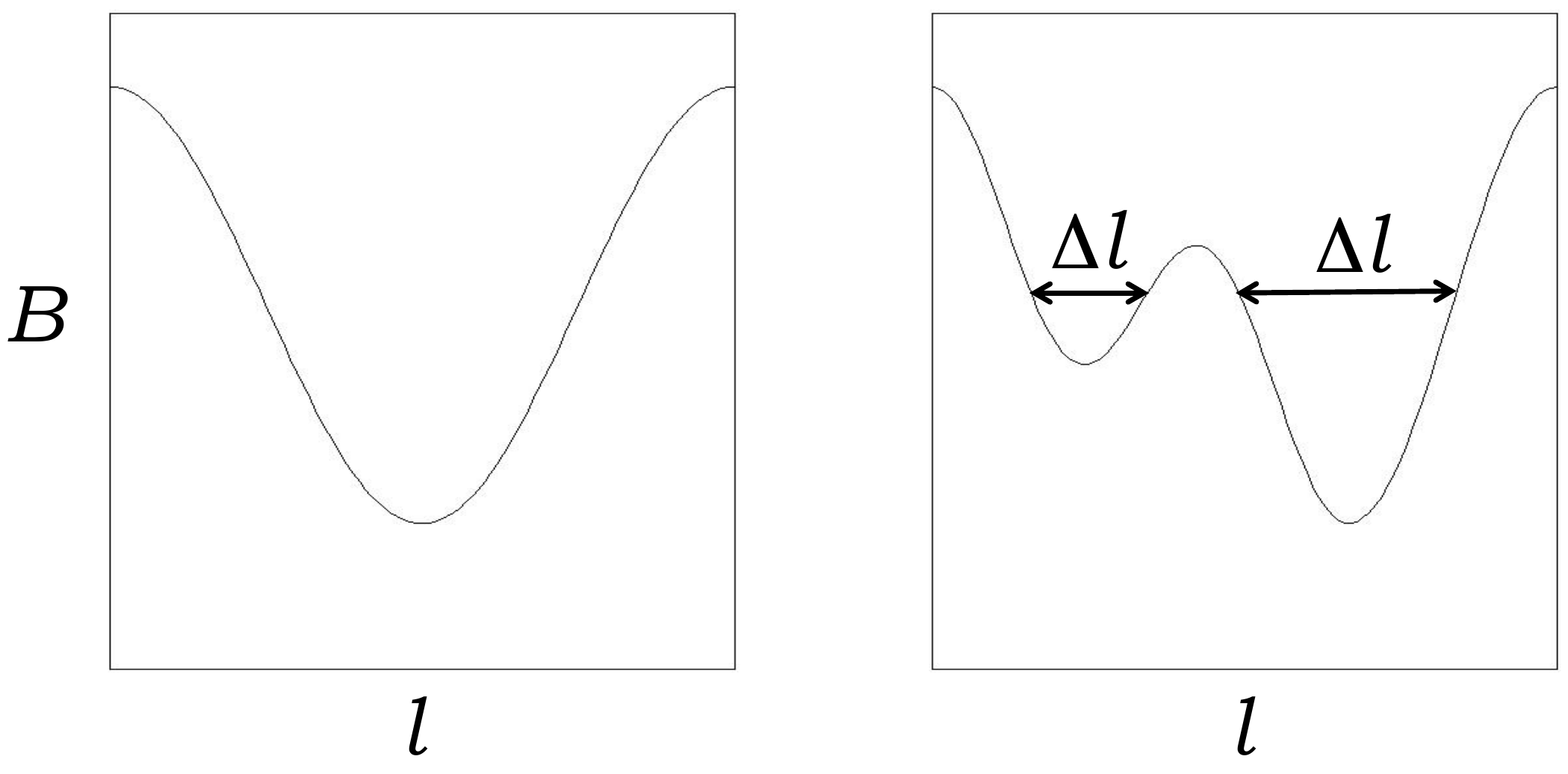}
\end{center}

\caption{\label{fig:carycondition} Profiles of $B(l)$ on a given magnetic field line (a) considered and (b) ignored in \cite{cary97a, cary97b}. Here $l$ is the arc length of the magnetic field line. In figure \ref{fig:carycondition}(b), the quantity $\Delta l (\rho, B)$, defined in \eq{eq:Deltaldef}, is sketched. Note that the definition of $\Delta l$ is independent of $\alpha$ in an omnigeneous stellarator, but it depends on the well. Even though the value of $B$ is the same, $\Delta l$ in the two contiguous wells in figure \ref{fig:carycondition}(b) is different.}
\end{figure}

Cary and Shasharina \cite{cary97a, cary97b} did not consider all possible classes of omnigenenous stellarators. References \cite{cary97a, cary97b} correctly show that the local minima of the magnetic field strength $B$ on a magnetic field line have the same value as the closest local minima in the contiguous magnetic field lines on the same flux surface. The same happens to local maxima. However, in addition to this condition, a large part of the discussion in references \cite{cary97a, cary97b} assumes that all these minima must have the same value on a given flux surface. Similarly, it is assumed that all the maxima must have the same value. In other words, it is assumed that there cannot be local minima larger than the global minimum, and there cannot be local maxima smaller than the global maximum (the situations that references \cite{cary97a, cary97b} consider and ignore are sketched in figure \ref{fig:carycondition}). However, it is not necessary for omnigeneity that all the minima in the flux surface have the same value and that all the maxima are the same, as we proceed to show by constructing an omnigeneous magnetic field that does not satisfy this condition. We first review the arguments given in \cite{cary97a, cary97b}, and we then use them to construct a solution with local minima and maxima of $B$ that do not have the same value.

Note that this article relies heavily on the work in \cite{cary97a, cary97b}, and only extends it. It does not detract from the importance of the original work. However, we believe that this correction is needed because the assumption that the value of all the local minima and maxima of the magnetic field in an omnigeneous stellarator only depends on the flux surface is considered to be true by a large part of the community.

%%%%%%%%%%%%%%%%%%%%%%%%%%%%%%%%%%%%%%%%%%%%%%%%%%%%%%%%%%%%%
\section{Conditions for an omnigeneous magnetic field} \label{sec:conditions}
%%%%%%%%%%%%%%%%%%%%%%%%%%%%%%%%%%%%%%%%%%%%%%%%%%%%%%%%%%%%%

Since to lowest order particles move only along magnetic field lines, it is convenient to use coordinates that clearly distinguish between motion across and along the magnetic field. We use a radial coordinate $\rho$ to label flux surfaces, an angle $\alpha$ to label magnetic field lines on a given flux surface (we give an exact definition of $\alpha$ below in \eq{eq:alphadef}), and the arc length of the magnetic field line to locate a particle along a magnetic field line once $\rho$ and $\alpha$ are given. To lowest order, the magnitude of the velocity $v$ and the pitch-angle-like variable $\lambda = v_\bot^2/v^2 B$ are constants of the motion, making the particle parallel velocity depend only on the magnitude of the magnetic field,
\begin{equation}
v_{||} = \sigma v \sqrt{1 - \lambda B(\rho, \alpha, l)},
\end{equation}
where $\sigma = \pm 1$ is the sign of the parallel velocity. Particles with $\lambda < B_\mathrm{max}^{-1}$, where $B_\mathrm{max} (\rho)$ is the maximum value of $B$ on the flux surface $\rho$, have a parallel velocity that never vanishes, and sample the entirety of the flux surfaces that the magnetic field lines cover ergodically (the number of flux surfaces in which magnetic field lines close on themselves is negligible). These passing particles always have a vanishing average radial magnetic drift. Particles with $\lambda > B_\mathrm{max}^{-1}$ are trapped between the bounce points $l_{b1} (\rho, \alpha, \lambda)$ and $l_{b2} (\rho, \alpha, \lambda)$ that satisfy $B(\rho, \alpha, l_{b1}) = \lambda^{-1} = B(\rho, \alpha, l_{b2})$. In general, trapped particles do not have vanishing average radial magnetic drift, and they can drift off flux surfaces. Because trapped particle orbits are periodic to lowest order, they must conserve the second adiabatic invariant
\begin{equation}
J_{||} (\rho, \alpha, v, \lambda) = \oint v_{||}\, \dd l = 2v \int_{l_{b1}}^{l_{b2}} \sqrt{1 - \lambda B(\rho, \alpha, l)}\, \dd l
\end{equation}
when they drift away from the magnetic field line where they started. These particles move to another flux surface if there is no trapped orbit in a contiguous magnetic field line on the same flux surface that has the same values of $v$, $\lambda$ and $J_{||}$. Thus, to make the radial drift of trapped particles vanish, we need to find magnetic field configurations in which
\begin{equation} \label{eq:dalphaJ}
\partial_\alpha J_{||} = 0.
\end{equation}
This condition, which is the definition of omnigeneity, must be satisfied for all $\lambda$ in the interval $[B_\mathrm{max}^{-1}, B_\mathrm{min}^{-1}]$, where $B_\mathrm{min} (\rho)$ is the minimum value of $B$ on the flux surface $\rho$. As a result, it imposes several important constraints on the magnetic field that were first deduced in \cite{cary97a, cary97b}. We proceed to discuss these constraints one by one.

\begin{enumerate}

\item \label{prop:Bmaxmin} Along the magnetic field line defined by $\rho$ and $\alpha$, the magnitude of the magnetic field $B$ has in general several local minima and maxima. We use $l_{m,j} (\rho, \alpha)$ and $B_{m, j} (\rho, \alpha)$ to denote the location and value of the $j$-th local minimum, and $l_{M, k} (\rho, \alpha)$ and $B_{M, k}(\rho, \alpha)$ for the location and the value $k$-th local maximum. We choose the indices $j$ and $k$ such that $B_\mathrm{min} (\rho) \leq B_{m, 1} (\rho, \alpha) \leq B_{m,2} (\rho, \alpha) \leq \ldots \leq B_{m, J} (\rho, \alpha) < B_\mathrm{max} (\rho)$ and $B_\mathrm{min} (\rho) < B_{M, 1} (\rho, \alpha) \leq B_{M,2} (\rho, \alpha) \leq \ldots \leq B_{M, K} (\rho, \alpha) \leq B_\mathrm{max} (\rho)$, where $J (\rho, \alpha)$ and $K (\rho, \alpha)$ are the number of local minima and maxima on the magnetic field line defined by $\rho$ and $\alpha$. Particles with $\lambda = B_{m,j}^{-1}$ located at $l_{m, j}$ do not move along magnetic field lines because they are completely trapped at $l = l_{m, j}$, but they move across magnetic field lines searching for a local minimum with value $B_{m,j}$. If there is no such a minimum within the flux surface defined by $\rho$, they need to move to one of the contiguous flux surfaces. Thus, to be omnigeneous, the magnetic field has to be such that the contour $B = B_{m, j}$ crosses all magnetic field line and closes on itself poloidally, toroidally or helically. This condition can be written as
\begin{equation}
\partial_\alpha B (\rho, \alpha, l_{m,j} (\rho, \alpha) ) = 0.
\end{equation}
Then, $B_{m,j}$ does not depend on $\alpha$, and we find
\begin{equation}
B_\mathrm{min} (\rho) \equiv B_{m,1} (\rho) \leq B_{m, 2} (\rho) \leq \ldots \leq B_{m, J} (\rho) < B_\mathrm{max} (\rho).
\end{equation}
However, in contrast to what is assumed in parts of \cite{cary97a, cary97b}, one cannot show that all the local minima are equal to $B_\mathrm{min} (\rho)$.

A similar but more sophisticated argument for particles with $\lambda$ values close to $B_{M,k}$ located at $l_{M,k}$ gives (see section III.B of \cite{cary97b})
\begin{equation}
\partial_\alpha B (\rho, \alpha, l_{M,k} (\rho, \alpha) ) = 0.
\end{equation}
As a result, the contour $B = B_{M,k}$ crosses all magnetic field lines and closes on itself poloidally, toroidally or helically. The local maxima $B_{M,k}$ do not depend on $\alpha$, leading to
\begin{equation}
B_\mathrm{min} (\rho) < B_{M,1} (\rho) \leq B_{M, 2} (\rho) \leq \ldots \leq B_{M, K} (\rho) \equiv B_\mathrm{max} (\rho).
\end{equation}
Again, in contrast to what is assumed in \cite{cary97a, cary97b}, it is not possible to prove that all the local maxima are the same as $B_\mathrm{max} (\rho)$.

\item \label{prop:Deltazeta} Condition \eq{eq:dalphaJ} can be used to show that the distance along a magnetic field line between two points $l_{b1} (\rho, \alpha, \lambda = B^{-1})$ and $l_{b2} (\rho, \alpha, \lambda = B^{-1})$ that have the same value of $B$ and are both within the same magnetic well is independent of $\alpha$ (see section III.C of \cite{cary97b}), that is,
\begin{equation} \label{eq:Deltaldef}
l_{b2} (\rho, \alpha, \lambda = B^{-1}) - l_{b1} (\rho, \alpha, \lambda = B^{-1}) = \Delta l (\rho, B).
\end{equation}
Note that the function $\Delta l$ depends on the well, as shown in figure \ref{fig:carycondition}(b). This condition can be easily restated in a particular type of straight field line coordinates known as Boozer coordinates \cite{boozer82}. We only need two properties of the Boozer coordinates. First, for $\theta$ the Boozer poloidal angle and $\zeta$ the Boozer toroidal angle, the angle $\alpha$ is
\begin{equation} \label{eq:alphadef}
\alpha = \theta - \iota \zeta,
\end{equation}
where $\iota(\rho)$ is the rotational transform. Second, for fixed $\rho$ and $\alpha$, the relation between the arc length of the magnetic field line and the toroidal angle, $\dd l/\dd \zeta$, only depends on $\rho$ and the value of $B$ at $(\rho, \alpha, l)$. This property of $\dd l/\dd \zeta$ and equation \eq{eq:Deltaldef} give that the angular separation between the two points $l_{b1}$ and $l_{b2}$ can only depend on $\rho$ and $B$,
\begin{equation} \label{eq:Deltazetadef}
\zeta (\rho, \alpha, l_{b2}) - \zeta (\rho, \alpha, l_{b1}) = \Delta \zeta ( \rho, B ).
\end{equation}

\item \label{prop:Bmaxstraight} Using result \eq{eq:Deltazetadef} and the fact that most flux surfaces are covered ergodically by a magnetic field line, it is possible to show that the contour $B = B_\mathrm{max}$ is a straight line in Boozer coordinates (see section III.C of \cite{cary97b}). Note that this is a property of the global maximum $B_\mathrm{max} (\rho)$, and not of the local maxima $B_{M,k}$.

\end{enumerate}

Importantly, due to the fact that the local maxima $B_{M, k}$ satisfy conditions \eq{prop:Bmaxmin} and \eq{prop:Deltazeta}, particle transitions from barely passing orbits to barely trapped orbits due to collisions or other effects are equivalent to the same transitions in a configuration with only one global maximum. In both cases, the radial magnetic drift vanishes at the maxima so transitioning particles do not take a large radial step, and barely trapped particles are radially confined. As a result, transitions from passing to trapped do not cause an increase in the radial flux of energy or particles.

\begin{figure}

\begin{center}
\includegraphics[width = 10 cm]{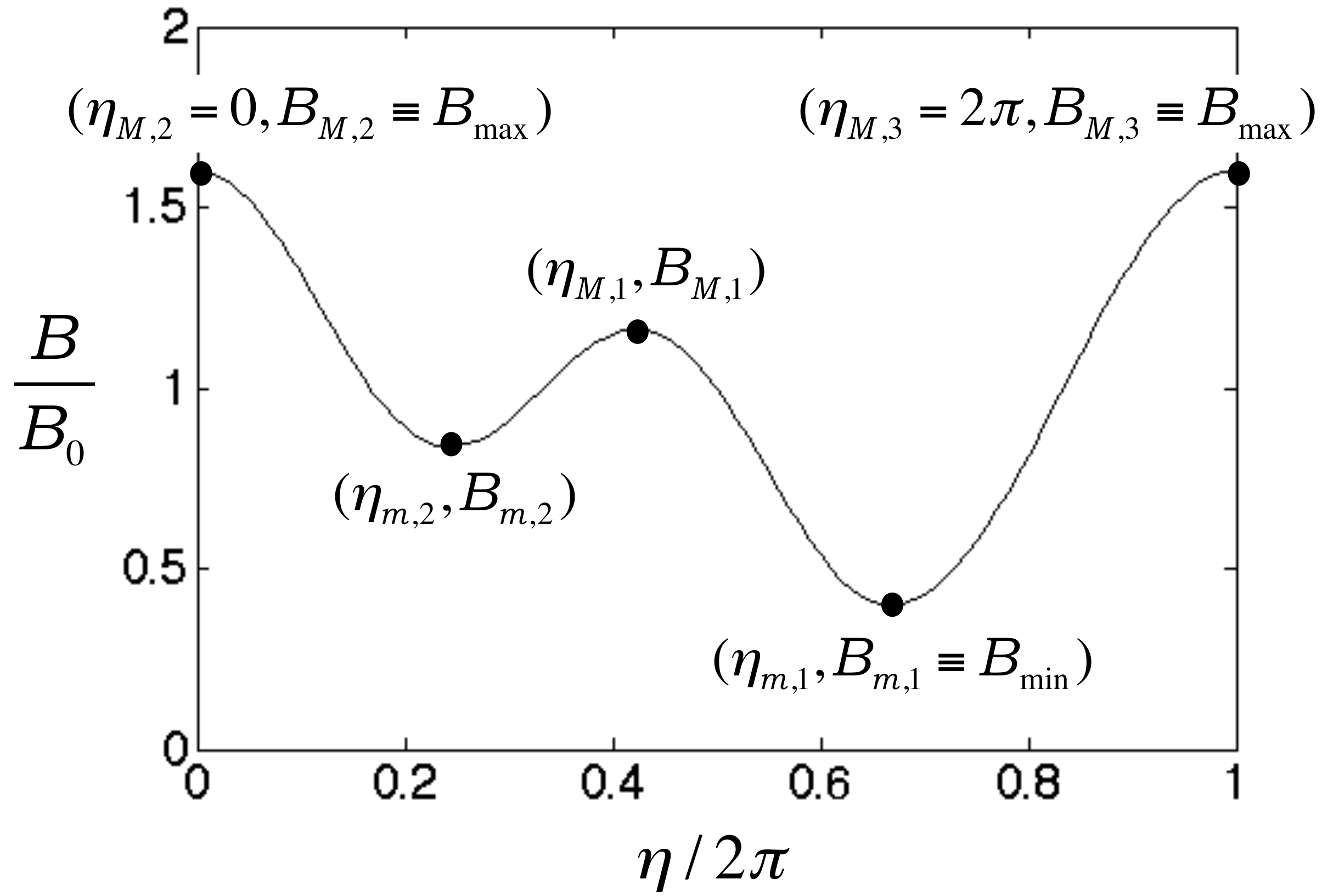}
\end{center}

\caption{\label{fig:B_eta} Function $B/B_0 = f (\eta)$ used as an example in this article. The local minima $B_{m,1}$ and $B_{m,2}$ and the local maxima $B_{M,1}$, $B_{M,2}$ and $B_{M,3}$ are given.}
\end{figure}

%%%%%%%%%%%%%%%%%%%%%%%%%%%%%%%%%%%%%%%%%%%%%%%%%%%%%%%%%%%%%
\section{Example}
%%%%%%%%%%%%%%%%%%%%%%%%%%%%%%%%%%%%%%%%%%%%%%%%%%%%%%%%%%%%%

The properties listed above can be used to construct an omnigeneous magnetic field with more than one local minima $B_{m,i}$ and more than one local maxima $B_{M,k}$. We follow a procedure similar to the one proposed in section V of \cite{cary97b}.

To define the function $B(\theta, \zeta)$, we use the intermediate coordinate $\eta (\theta, \zeta)$, defined such that the contours of constant $\eta$ are contours of constant $B$,
\begin{equation}
\frac{B}{B_0} = f (\eta),
\end{equation}
where $B_0$ is a normalization constant. The function $f(\eta)$ is the function that gives the number and type of local minima and maxima. To give an example with local minima and maxima that are different from $B_\mathrm{min}$ and $B_\mathrm{max}$, we focus on the function
\begin{equation} \label{eq:fexample}
f(\eta) = \left \{
\begin{array}{l l}
1 + 0.3 [ \cos (3\eta/4) + \cos(9\eta/4)]& \mathrm{for}\quad \eta \in [0, 4\pi/3], \\
1 + 0.6 \cos(\pi - 3\eta/2) & \mathrm{for}\quad \eta \in [4\pi/3, 2\pi],
\end{array}
\right .
\end{equation}
plotted in figure \ref{fig:B_eta}, with two local minima, $B_{m, 1} \equiv B_\mathrm{min} = 0.4B_0$ and $B_{m, 2} = 0.837B_0$, located at $\eta_{m, 1} = 4\pi/3$ and $\eta_{m,2} = 0.488\pi$, and three local maxima, $B_{M,1} = 1.163B_0$ and $B_{M,2} \equiv B_{M,3} \equiv B_\mathrm{max} = 1.6B_0$, located at $\eta_{M,1} = 0.845\pi$, $\eta_{M,2} = 0$ and $\eta_{M,3} = 2\pi$.

\begin{figure}

\begin{center}
\includegraphics[width = 13 cm]{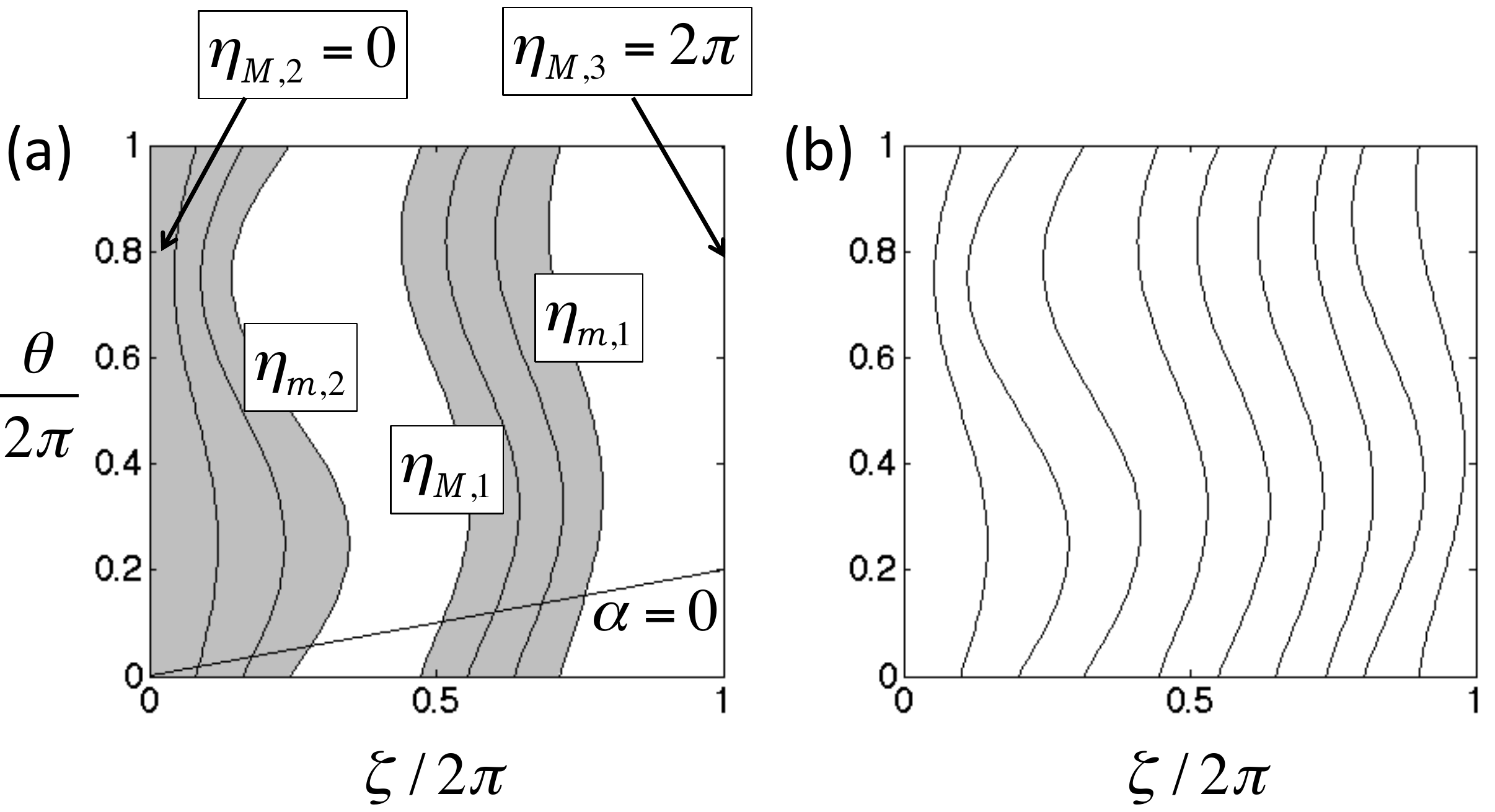}
\end{center}

\caption{\label{fig:problemset} (a) Sketch of the information needed to determine completely the function $\eta(\theta, \zeta)$. The function $\eta(\theta, \zeta)$ must be known in the shaded areas and on the line $\alpha = 0$. (b) Contours of the function $\eta(\theta, \zeta)$.}
\end{figure}

The function $\eta (\theta, \zeta)$ is chosen so that $B(\eta(\theta, \zeta))$ satisfies properties \eq{prop:Bmaxmin}, \eq{prop:Deltazeta} and \eq{prop:Bmaxstraight} in section~\ref{sec:conditions}. We start by choosing the shape of the contours $\eta = \eta_{M,2}$ and $\eta = \eta_{M,3}$ that correspond to $B_\mathrm{max}$. These contours must satisfy property \eq{prop:Bmaxstraight}. Without loss of generality, we choose the contour $B = B_\mathrm{max}$ to be $\zeta = 0, 2\pi$. Then, $\eta = \zeta$ at $\zeta = 0, 2\pi$ (see the sketch in figure \ref{fig:problemset}(a)). According to property \eq{prop:Bmaxmin}, the contours $\eta \in \{\eta_{m,1}, \eta_{m,2}, \eta_{M,1}, \eta_{M,2}, \eta_{M,3}\}$ must close on themselves. Since we have already chosen the contours $\eta = \eta_{M,2}, \eta_{M,3}$ to be $\zeta = 0, 2\pi$, and contours cannot cross each other, the contours $\eta = \eta_{m,1}, \eta_{m,2}, \eta_{M,1}$, and consequently all the contours of constant $B$, must close poloidally. Thus, the function $\eta$ is defined by
\begin{equation} \label{eq:zetaeta}
\zeta = \eta + G(\theta, \eta),
\end{equation}
where the function $G$ vanishes for $\eta = 0, 2\pi$ and is defined such that the relation between $\zeta$ and $\eta$ is invertible.

To completely determine $G(\theta, \eta)$ we only need to know this function in certain regions of the $(\theta, \eta)$ plane. In particular, for the function $B/B_0 = f(\eta)$ in equation \eq{eq:fexample} and figure \ref{fig:B_eta}, we need information in the regions highlighted in figure \ref{fig:problemset}(a). We need to know $G(\theta, \eta)$
\begin{itemize}

\item in the region $\eta \in [0, \eta_{m,2}]$, where
\begin{equation} \label{eq:G0etam2}
G(\theta, \eta) = g (\theta, \eta);
\end{equation}

\item partially, in the region $\eta \in [\eta_{M,1}, \eta_{m,1}]$, where the difference
\begin{equation} \label{eq:GetaM1etam1}
G(\theta, \eta) - G(\theta, \eta_{M,1}) = h (\theta, \eta)
\end{equation}
must be given;

\item and on one magnetic field line (for example, $\alpha = 0$), where
\begin{equation} \label{eq:alphaequal0}
G(\theta_{\alpha=0} (\eta), \eta) = y (\eta).
\end{equation}
Here $\theta_{\alpha=0} (\eta)$ is the solution to equations \eq{eq:zetaeta} and $\alpha = \theta - \iota \zeta = 0$ for a given $\eta$.

\end{itemize}

\begin{figure}

\begin{center}
\includegraphics[width = 10 cm]{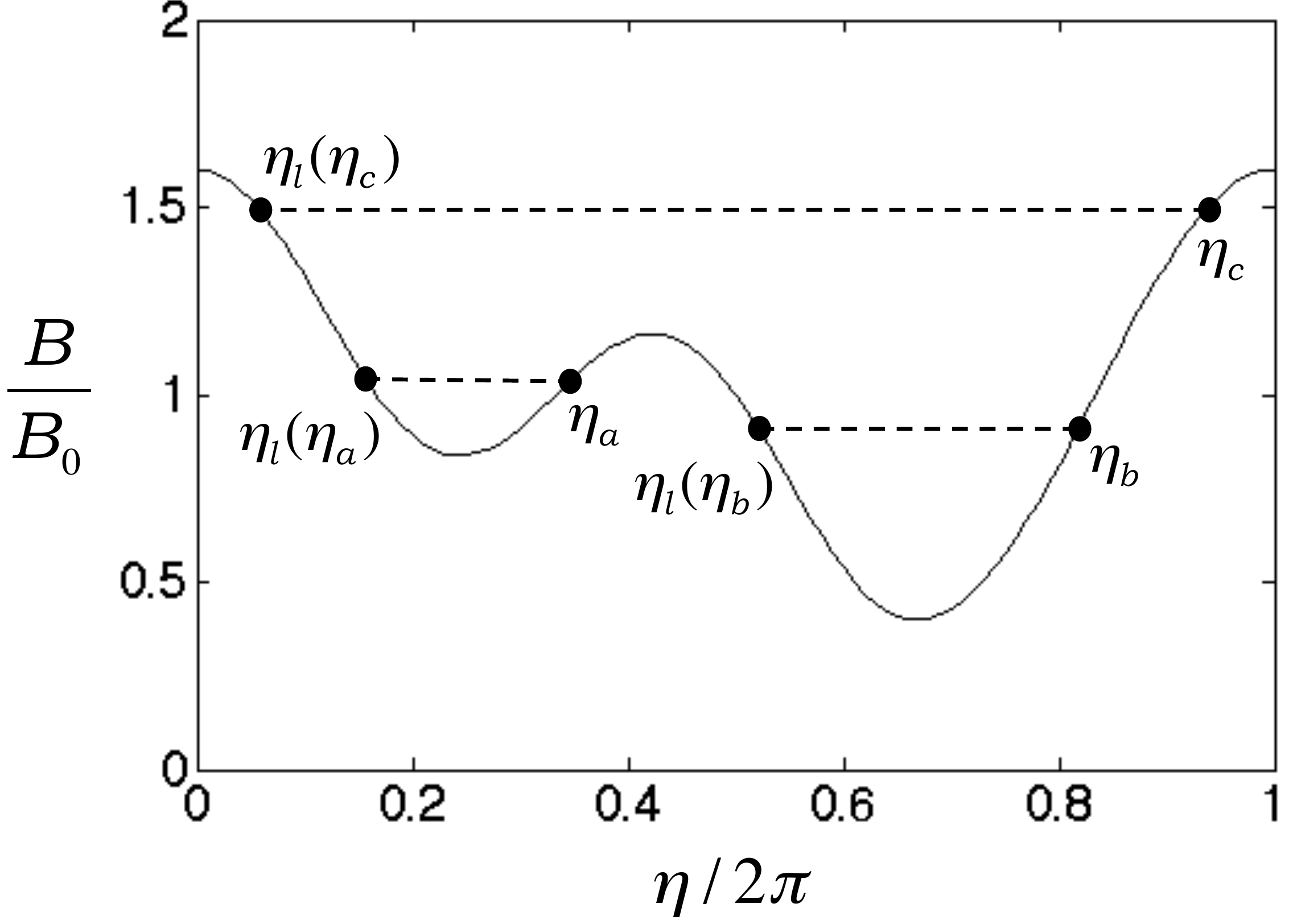}
\end{center}

\caption{\label{fig:eta_l} Function $\eta_l (\eta)$ evaluated for three different values of $\eta$: $\eta_a$, $\eta_b$ and $\eta_c$.}
\end{figure}

Using equation \eq{eq:alphaequal0} we can obtain the angular difference $\Delta \zeta (B)$ that according to property \eq{prop:Deltazeta} can only depend on the value of the magnitude of the magnetic field. Since $\eta$ and $B$ are almost equivalent, we write $\Delta \zeta$ as a function of $\eta$. All the values of $B$ can be mapped to the intervals $\eta \in [\eta_{m, 2}, \eta_{M,1}]$ and $\eta \in [\eta_{m,1}, 2\pi]$. In these intervals, we define the function
\begin{equation}
\Delta \zeta (\eta) = \eta - \eta_l(\eta) + y (\eta ) - y (\eta_l (\eta)).
\end{equation}
The function $\eta_l (\eta)$ gives the location $\eta_l$ to the immediate left of $\eta$ that satisfies $f(\eta_l) = f(\eta)$. The function $\eta_l$ is sketched in figure \ref{fig:eta_l}.

Once the function $\Delta \zeta (\eta)$ is known, we can use the fact that $\Delta \zeta$ only depends on the magnitude $B$ to calculate $G(\theta, \eta)$ from \eq{eq:G0etam2} and \eq{eq:GetaM1etam1}. We find
\begin{equation} \label{eq:finalG}
\fl G(\theta, \eta) = \left \{
\begin{array}{l l}
g (\theta, \eta) & \mathrm{for}\quad \eta \in [0, \eta_{m,2}], \\
\Delta \zeta (\eta) - \eta + \eta_l(\eta) + g(\theta - \iota \Delta \zeta(\eta), \eta_l(\eta)) & \mathrm{for}\quad \eta \in [\eta_{m,2}, \eta_{M,1}], \\
h (\theta, \eta) + G(\theta, \eta_{M,1}) & \mathrm{for} \quad \eta \in [\eta_{M,1}, \eta_{m,1}], \\
\Delta \zeta (\eta) - \eta + \eta_l(\eta) + G(\theta - \iota \Delta \zeta(\eta), \eta_l(\eta)) & \mathrm{for}\quad \eta \in [\eta_{m,1}, 2\pi].
\end{array}
\right .
\end{equation}
Here
\begin{equation}
\fl G(\theta, \eta_{M,1}) = \Delta \zeta (\eta_{M,1}) - \eta_{M,1} + \eta_l(\eta_{M,1}) + g(\theta - \iota \Delta \zeta(\eta_{M,1}), \eta_l(\eta_{M,1})).
\end{equation}
Note that the function $G(\theta - \iota \Delta \zeta (\eta), \eta_l (\eta))$ that appears in the formula for $G(\theta, \eta)$ for $\eta \in [\eta_{m,1}, 2\pi]$ is known because for $\eta \in [\eta_{m,1}, 2\pi]$, $\eta_l (\eta) \in [0, \eta_{m,2}] \cup [\eta_{M,1}, \eta_{m,1}]$.

\begin{figure}

\begin{center}
\includegraphics[width = 10 cm]{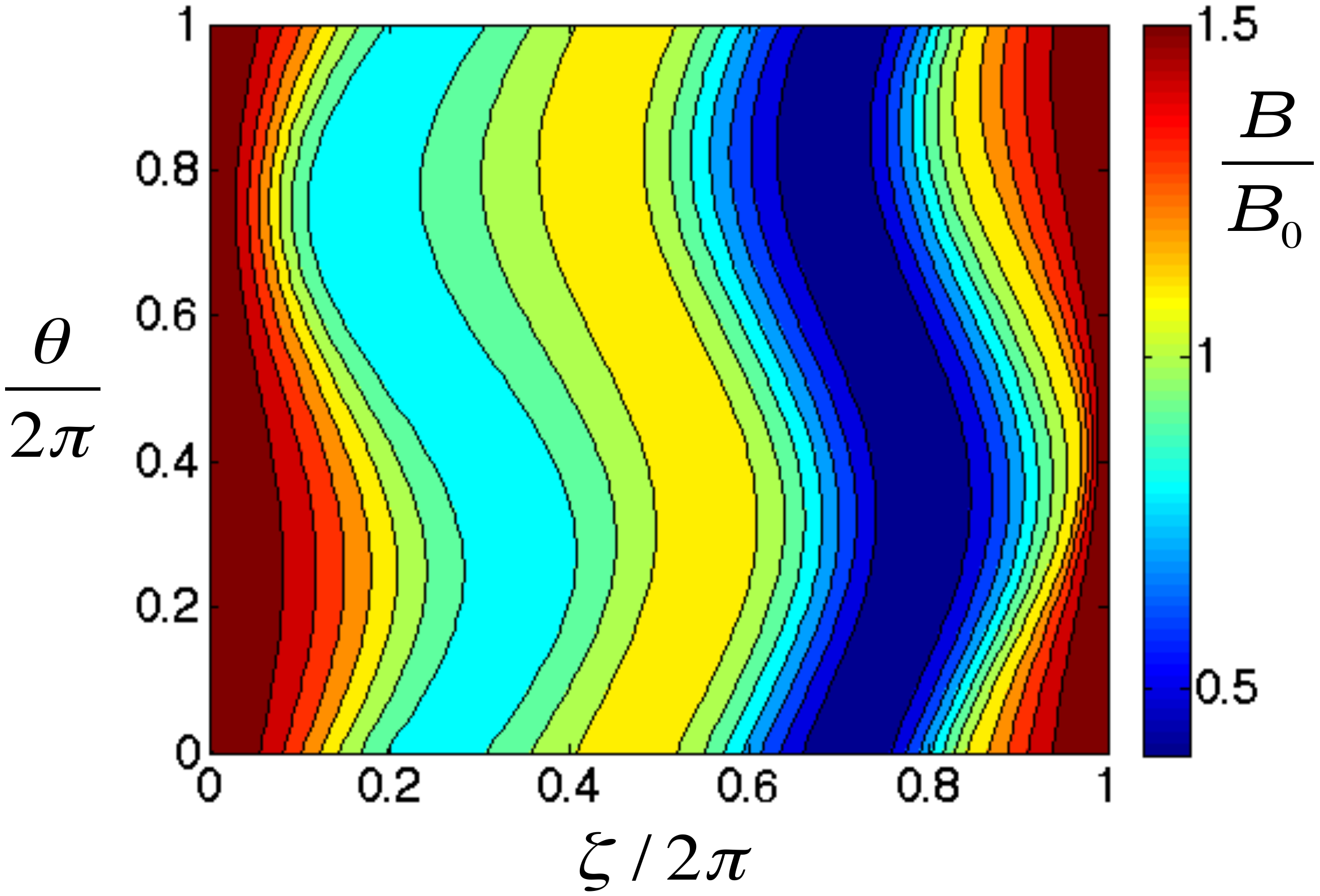}
\end{center}

\caption{\label{fig:Bomni} Contour plot of the omnigeneous magnetic field.}
\end{figure}

Using equation \eq{eq:finalG} in a flux surface with rotational transform $\iota = 0.2$ and with the functions
\begin{equation}
g(\theta, \eta) = 0.3\pi \sin \theta \sin (\eta/2),
\end{equation}
\begin{equation}
h(\theta, \eta) = 0.3\pi\sin \theta [\sin (\eta/2) - \sin(\eta_{M,1}/2)]
\end{equation}
and
\begin{equation}
y(\eta) = 0.3\pi \sin (\theta_{\alpha = 0}(\eta)) \sin (\eta/2),
\end{equation}
we obtain the function $\eta(\theta, \zeta)$ given in figure \ref{fig:problemset}(b). The final omnigeneous magnetic field $B / B_0 =  f(\eta(\theta, \zeta))$ is given in figure \ref{fig:Bomni}.

%%%%%%%%%%%%%%%%%%%%%%%%%%%%%%%%%%%%%%%%%%%%%%%%%%%%%%%%%%%%%
\section{Discussion}
%%%%%%%%%%%%%%%%%%%%%%%%%%%%%%%%%%%%%%%%%%%%%%%%%%%%%%%%%%%%%

With the example in figure \ref{fig:Bomni}, we have shown that in an omnigeneous stellarator, the values of the magnetic field strength at local minima do not have to be the same on a given flux surface. The same is true for the local maxima. 

An omnigenenous toroidal flux surface must satisfy several conditions. The local maxima (minima) on a magnetic field line must have the same value as the closest maxima (minima) in the contiguous magnetic field line, forming curves of constant magnetic field strength that must close poloidally, toroidally or helically. In addition to this condition, the contours corresponding to the global maximum of the magnetic field strength on the flux surface have to be straight lines in Boozer coordinates. A consequence of these conditions is that the magnetic field strength wells extend until they close on themselves poloidally, toroidally or helically. The distance along a field line between two contours with the same value of magnetic field strength on two opposite sides of a well is a function only of the flux surface and of the well.

The extra assumption that all the local minima and all the local maxima have the same value has been used for simplicity in previous work, such as \cite{helander09, landreman12}. The results of these papers only need to be generalized slightly to account for local minima and maxima with values different from the global minimum and maximum on the flux surface, but the qualitative results are probably unchanged.

In certain classes of stellarators, it may be beneficial for the minima of the magnetic field strength on a flux surface to have similar values \cite{mynick82}, but we have shown that in general this is not a necessary condition for optimized stellarators. The more general conditions for omnigeneity discussed in this article ensure that the neoclassical fluxes do not scale inversely with collisionality. These conditions have been derived for a flux surface without consideration of the neighboring flux surfaces because we have limited our analysis to particles with small radial orbit widths. The study of long term confinement of energetic particles \cite{lotz92, drevlak14} requires a more careful analysis than the one performed here, but one would expect that optimizing neoclassical fluxes would improve energetic particle confinement.

In practice, the design of an omnigeneous stellarator experiment would be based upon the multiple-criteria optimization procedures analogous to those used in references \cite{gori99, subbotin06, kasilov13} to identify a 3D MHD equilibrium. As part of such a design, collisionless charged particle losses would need to be computed directly using the codes based on the guiding center drift equations, as in \cite{lotz92, drevlak14}. However, due to the high dimensionality of the optimization problem, simple but robust optimization criteria are required to use the optimization codes effectively, for selecting appropriate weighted cost functions, initial configurations, and search algorithms. Our results here give such criteria. An optimization that imposed that the local minima and maxima must be the same on a given flux surface could give a stellarator close to omnigeneity, but it would have ignored a large part of the allowed parameter space, therefore missing potentially better solutions.

%%%%%%%%%%%%%%%%%%%%%%%%%%%%%%%%%%%%%%%%%%%%%%%%%%%%%%%%%%%%%
\section*{Acknowledgments}
%%%%%%%%%%%%%%%%%%%%%%%%%%%%%%%%%%%%%%%%%%%%%%%%%%%%%%%%%%%%%

The authors would like to thank J.R. Cary and S.G. Shasharina for helpful discussions. This work has been carried out within the framework of the EUROfusion Consortium and has received funding from the European Union's Horizon 2020 research and innovation programme under grant agreement number 633053. The views and opinions expressed herein do not necessarily reflect those of the European Commission. This research was supported in part by the RCUK Energy Programme (grant number EP/I501045) and by grant ENE2012-30832, Ministerio de Econom\'{\i}a y Competitividad, Spain.

%%%%%%%%%%%%%%%%%%%%%%%%%%%%%%%%%%%%%%%%%%%%%%%%%%%%%%%%%%%%%%
\section*{References}
%%%%%%%%%%%%%%%%%%%%%%%%%%%%%%%%%%%%%%%%%%%%%%%%%%%%%%%%%%%%%%

\end{document}